\documentclass{aa}
\usepackage{txfonts}
\usepackage{graphicx}
\usepackage{natbib}

\begin{document}

\title{The dust condensation sequence in red super-giant stars}

\author{T. Verhoelst \inst{1,2} \fnmsep
        \thanks{Postdoctoral Fellow of the Fund for Scientific Research,
                Flanders}
  \and N. Van der Zypen \inst{1}
  \and S. Hony \inst{3}
  \and L. Decin \inst{1,\star}
  \and J. Cami \inst{4}
  \and K. Eriksson \inst{5}
}

\offprints{T. Verhoelst, \email{tijl.verhoelst@ster.kuleuven.be}}

\institute{Instituut voor Sterrenkunde, K.U. Leuven, Celestijnenlaan
  200D, B-3001 Leuven, Belgium
  \and University of Manchester, Jodrell Bank Centre for Astrophysics, Manchester, M13 9PL, U.K.
  \and Laboratoire AIM, CEA/DSM - CNRS - Université Paris Diderot,
DAPNIA/SAp, 91191 Gif sur Yvette, France
  \and Physics and Astronomy Dept, University of Western Ontario,
  London ON N6A 3K7, Canada
  \and  Institute for Astronomy and Space Physics, Box 515, 75120
Uppsala, Sweden  
}

\date{Received  / Accepted }

\abstract{Red super-giant (RSG) stars exhibit significant mass loss
through a slow and dense wind. They are often considered to be the
more massive counter-parts of Asymptotic Giant Branch (AGB)
stars. While the AGB mass-loss is linked to their strong pulsations,
the RSG are often only weakly variable. This raises the question
whether their wind-driving mechanism and the dust composition in the
wind are the same.}{To study the
conditions at the base of the wind, by determining the dust
composition in a sample of RSG. The dust composition is thought to be
sensitive to the density, temperature and acceleration at the base
of the wind. To compare the derived composition with the composition
found in AGB star winds.}{We compile a sample of 27 RSG infrared
spectra (ISO-SWS) and supplement these with photometric measurements
to obtain the full spectral energy distribution (SED). These data are
modelled using a dust radiative transfer code, taking into account the
optical properties of relevant candidate materials. The results are
scrutinised for correlations in terms of mass-loss rate, density at
the inner edge of the dust shell and stellar parameters.}{ We find
(1) strong correlations between dust composition, mass-loss rate and
stellar luminosity, roughly in agreement with the theoretical dust
condensation sequence, (2) the need for a continuous (near-)IR dust
opacity and tentatively propose amorphous carbon, and (3) significant
differences with AGB star winds: presence of PAHs, absence of 'the'
13~$\mu$m band, and a lack of strong water bands.}{ Dust
condensation in RSG is found to experience a similar freeze-out
process as in AGB stars. Together with the positive effect of the
stellar luminosity on the mass-loss rate, this suggests that radiation
pressure on dust grains is an important ingredient in the driving
mechanism.  Still, differences with AGB stars are manifold and thus
the winds of RSG deserve separate studies.}



\maketitle

\section{Introduction}

As stars of initial mass $8\le$\,M$_{\rm{init}}\le 40\,\rm{M}_{\odot}$
evolve off the main sequence and reach the core helium-burning phase,
they turn into red super-giant (RSG) stars, the largest and most
luminous of all stars \citep[for a recent review on massive star
  observation and evolution, see][]{Massey2003}. Through their stellar wind
and eventually when they end their life through a supernova explosion,
they enrich the ISM with heavy elements and large amounts of kinetic
energy, possibly triggering new star formation
\citep{Gehrz1989,Herbst1977}.The dusty stellar winds of RSG not only
influence their evolution, but also complicate the determination of
general properties such as the effective temperature and surface
gravity, e.g. through reddening which appears different from that
through the ISM \citep{Massey2005}.

Most RSG are irregular variables with smaller amplitudes and higher
effective temperatures than AGB stars. Moreover, they often show
significant chromospheric activity.  These properties make them
intrinsically very different from their lower-mass counterparts in
which strong pulsations and cool molecular layers are believed to be
crucial ingredients for the mass loss \citep[][and further
updates]{Hoefner1998}. Instead, convection, rotation and/or
chromospheric activity may turn out to play an important role. For
example, \cite{Josselin2007} find a correlation between mass-loss rate
and the strength of the photospheric turbulence.

While the mechanism responsible for the initial levitation of the
atmosphere may therefore be different in RSG as compared to
AGB stars, both types of stars are often used together in studies of
the dust condensation sequence and wind driving mechanism
\citep[e.g. ][]{Matsuura2005,Dijkstra2005}. Nevertheless,
\cite{Speck2000} find differences in dust composition between RSG and
AGB stars, with the former often showing Ca-Al-rich silicates instead
of magnesium silicates. 

Assuming a dust condensation sequence such as that presented by
\cite{Tielens1990} and placing it in an expanding stellar wind, one
expects a process called ``freeze-out'', where the sequence is not
completed but stops at an intermediate product when the wind density
drops below the density required for the next step. This freeze-out is
now tentatively observed in AGB stars \citep[e.g.][]{Heras2005}. If
the dust condensation and the wind dynamics are similar for the RSG, one
may also expect to observe incomplete dust condensation.  We test this
hypothesis through the analysis of all ISO-SWS spectra of
RSG.

In Sect.\,\ref{sec:sample}, we present the construction of the sample
and its main characteristics. Sect.\,\ref{sec:obs} deals with the
reduction of the ISO-SWS spectra. In the following section, we discuss
the details of our modelling strategy, including the photosphere
models, dust components, radiative transfer method and fitting
strategy. In Sect.\,\ref{sec:sum}, we analyse an ``averaged'' RSG dust
spectrum to check for any major short comings in our modelling
approach. The results on all individual sample stars are then
summarised in Sect.\,\ref{sec:results} and trends and correlations
between the derived parameters are presented. The discrepancies
between models and observed spectra are studied in
Sect.\,\ref{sec:residuals}. We end with a discussion
 and the conclusions (Sect.\,\ref{sec:discussionandconclusions}).

\section{The sample}
\label{sec:sample}

The availability of an ISO-SWS spectrum is an essential ingredient for
our research aims and the first step is therefore the selection of all
stars with spectral types M and K and luminosity classes I and II from
the list of usable 2.4--45\,$\mu$m ISO-SWS spectra presented by
\cite{Sloan2003}. From this list of 41\,stars, we remove the RV\,Tauri
star R\,Sct as it is a Post-AGB object. The spectra of XX\,Per,
IRC\,+60370 and WX Cas show technical anomalies and are also removed
from our list. Distances are derived from the {\sc hipparcos} parallax
measurement \citep{Perryman1997} or from the assumed distances to the
clusters containing our program stars \citep{Humphreys1978}. For
IRC+40427, no reliable distance estimate is available. We assume it to
be a typical 2\,kpc, which may in fact be to nearby, as the observed
reddening is much stronger than the combined effect of that distance
and its circumstellar shell (see Sect.\ref{sec:results}). As the
strong reddening may also be a consequence of a non-spherical
distribution of the circumstellar dust, we refrain from deriving a distance from the
observed reddening. 

Effective temperatures and bolometric corrections are estimated from
the spectral type based on the relations found by \cite{Levesque2005}.
Comparison with the evolutionary tracks of \cite{Lejeune2001} reveals
24 bona fide red supergiants, and 3 strong candidates\footnote{Stars
removed from the list because they are either on the AGB or even on
the RGB: T\,Cet, WX\,Cas, BD+59\,594, $\iota$ Aur, HD\,90586, RT\,Car,
BC\,Cyg, CIT\,11 and IRC+60370, $\beta$\,Cap, S\,Pav and
$\alpha$\,UMa}. VY\,CMa is an RSG experiencing extreme mass loss
resulting in an optically thick circumstellar environment (CSE). The
detailed analysis required to derive reliable dust composition
parameters for VY\,CMa is beyond the scope of this paper, and we refer
to \cite{Harwit2001} for a study of its ISO-SWS spectrum.  

We use these 26 stars for our study. Their designations and main
characteristics are presented in Table\,\ref{tab:sample}. The overlap
with the samples of \cite{Sylvester1994}, \cite{Speck2000} and
\cite{Levesque2005} is 8,11 and 14 stars, respectively.  

\cite{Kraemer2002} present a classification of, amongst others, our
sample stars on the basis of their photospheric and dust
characteristics. Most stars are in the {\it 2.SEa, 2.SEb} or {\it 2.SEc}
classes, meaning they show strong dust features from silicates and
alumina, and molecular bands in the photospheric part of the
spectrum. 

\begin{table*}
 \begin{center}
 \caption{
  Object designations and general characteristics. Star with their
  names printed in italic have luminosities which also allow an AGB
  star identification. The bolometric magnitudes were computed from
  the dereddened K band magnitudes, using the bolometric corrections of
  \cite{Levesque2005}.
   \label{tab:sample}
  }
 \vspace{1ex}
 \begin{tabular}{lcccccccc}
  \hline
  \hline
  Source & ISO TDT  & Spectral & Variability$^{\rm{a}}$ & Period$^{\rm{a}}$ & Kraemer$^{\rm{b}}$ &  Distance$^{\rm{c}}$ & A$_{\rm{V}}$$^{\rm{d}}$ & M$_{\rm{bol}}$ \\
         & number   & Type     & Type        & [d]        & class       &  [pc]     &            &  \\
  \hline
  HD 14242       & 61301202 & M2Iab      & Lc   &      & 2.SEa:  & 2290 & 1.63 & -6.93\\
  AD PER         & 78800921 & M2.5Iab    & SRc  & 363  & 2.SEa:  & 2290 & 1.63 & -7.42\\
  HD 14404       & 45501704 & M1Iab      & Lc   &      & 2.SEap: & 2290 & 1.63 & -7.23\\
  SU PER         & 43306303 & M3Iab      & SRc  & 533  & 2.SEc   & 2290 & 1.63 & -7.90\\
  RS PER         & 45501805 & M4Iab      & SRc  & 245  & 2.SEc   & 2290 & 1.63 & -7.74\\
  S PER          & 42500605 & M4.5Iab    & SRc  & 822  & 3.SE    & 2290 & 1.63 & -8.18\\
  HD 14580       & 42701401 & M1Iab      & Lc   &      & 1.NO:   & 2290 & 1.63 & -6.49\\
  HD 14826       & 61601203 & M2Iab      & Lc   &      & 2.SEa:  & 2290 & 1.63 & -7.36\\
  YZ PER         & 47301604 & M2Iab      & SRb  & 378  & 2.SEc   & 2290 & 1.63 & -7.50\\
  W PER          & 63702662 & M4.5Iab    & SRc  & 485  & 2.SEc   & 2290 & 1.63 & -7.74\\
  {\it RHO PER}  & 79501105 & M4II       & SRb  & 50   & 1.NO    & 100  & 0.82 & -4.20\\
  {\it HR 1939}  & 86603434 & M2Iab      & Lc   &      & 2.SEa   & 420  & 0.66 & -4.51\\
  ALF ORI        & 69201980 & M2Iab      & SRc  & 2335 & 2.SEcp  & 131  & 0.09 & -7.19\\
  {\it HD 90586} & 25400410 & M2Iab/Ib   &      &      & 2.SEc   & 641  & 0.40 & -4.76\\
  R CEN          & 07903010 & M5IIevar   & M    & 546  & 2.SEap  & 641  & 0.95 & -6.94\\
  ALF SCO        & 08200369 & M1.5Iabb   & Lc   &      & 2.SEcp  & 185  & 0.72 & -7.86\\
  ALF HER        & 28101115 & M5Iab      & SRc  &      & 1.NOp   & 117  & 0.42 & -5.99\\
  {\it SIG OPH}  & 10200835 & K3Iab      &      &      & 1.NO    & 360  & 0.68 & -4.33\\
  HR 7475        & 31601515 & K4Ib       & N:   &      & 1.NO    & 704  & 0.91 & -5.87\\
  NR VUL         & 53701751 & K3Iab      & Lc   &      & 2.SEc   & 2000 & 2.98 & -8.01\\
  BD+35 4077     & 73000622 & M2.5Iab    & Lc   &      & 2.SEb   & 1820 & 2.61 & -7.48\\
  RW CYG         & 12701432 & M3Iab      & SRc  & 550  & 2.SEc   & 1200 & 3.54 & -7.77\\
  IRC +40427     & 53000406 & M1:Iab     &      &      & 2.SEap: & 2000 & 1.79 & -6.89\\
  MU CEP         & 08001274 & M2Ia       & SRc  & 730  & 2.SEc   & 830  & 1.84 & -8.86\\
  V354 CEP       & 41300101 & M2.5Iab    & Lc   &      & 2.SEc   & 3500 & 2.00 & -8.51\\
  U LAC          & 41400406 & M4Iab:e    & SRc  &      & 2.SEc   & 3470 & 1.74 & -8.30\\
  PZ CAS         & 09502846 & M3Iab      & SRc  & 925  & 2.SEc   & 2510 & 2.11 & -8.89\\

\hline
 \end{tabular}		  	      
\begin{list}{}{}
\item[$^{\mathrm{a}}$] Combined General Catalogue of Variable Stars \citep{Kholopov1998}.  
\item[$^{\mathrm{b}}$] \cite{Kraemer2002}
\item[$^{\mathrm{c}}$] Distances below 1\,kpc come from parallax measurements with {\sc hipparcos} \citep{Perryman1997}. Distances equal to 2 kpc are actually unknown and distances greater than 1\,kpc are cluster distances from \cite{Humphreys1978}. $\mu$\,Cep is an exception: we use the distance of \cite{Humphreys1978} instead of the poor parallax measurement.
\item[$^{\mathrm{d}}$] ISM extinction estimates for stars within 1\,kpc come from the galactic model of \cite{Arenou1992}. All other stars and $\mu$\,Cep have estimates based on the extinction toward early-type stars in the corresponding clusters. 
\end{list}
 \end{center}
\end{table*}

\section{The observations}
\label{sec:obs}
The Infrared Space Observatory Short Wavelength Spectrometer
\citep[ISO-SWS][]{Kessler1996} observations were done in the AOT1
observing mode, resulting in a low-resolution full grating scan,
except for S\,Per, which was observed at full instrumental resolution
in the AOT06 mode. The observations were then processed using the SWS
interactive analysis product, IA \citep[see][]{deGraauw} using
calibration files and procedures equivalent to pipeline version
10.1. Further data processing consisted of extensive bad data removal
and rebinning on a fixed resolution ($\lambda$/$\Delta\lambda$=200)
wavelength grid. The bad data-removal uses the large amount of
redundancy in the available spectral scans to identify cosmic hits,
sudden changes in dark-current and other artifacts that cause the
signal of a single or multiple detectors to diverge from the mean of
the other spectral scans. In order to combine the different subband
into one continuous spectrum from 2 to 45 $\mu$m we have applied
scaling factors or offsets. In general the match between the different
subbands is good and the applied scaling/offsets are small compared to
the flux calibration uncertainties with a few exceptions: Alpha Her
and NR Vul show much larger differences between the different subbands
than can be expected on the basis of the flux-calibration
uncertainties alone. This is most likely due to mispointed
observations. For these two observations we use the IRAS point-source
fluxes as a guideline for splicing the spectrum. The RMS noise
increases with increasing wavelength, in particular at $\lambda$ $>$
27~$\mu$m. The flux levels of our sample stars drop with wavelength,
in particular if there is little on no dust excess. This causes some
of the longest wavelength parts of the spectra to be noise
dominated. See as the worst example H\,14580 in
Fig.\,\ref{fig:fits}. The noise-dominated parts are not used in the
analysis.

For a proper estimation of the photospheric irradiation and underlying
continuum, we prefer to
construct a spectral energy distribution (SED) covering wavelengths
from the UV to the far-IR. 
Optical photometry can be found in
\cite{Johnson1966a}, \cite{Johnson1966b}, \cite{Mendoza1967},
\cite{Lee1970}, \cite{Cousins1971}, \cite{Humphreys1974},
\cite{Wawrukiewicz1974}, \cite{Nicolet1978} and
\cite{Kharchenko2001}. Near-IR photometry is available for all our
sources from the 2MASS catalogue \citep{Skrutskie2006} and the far-IR
side of the SED can be covered with the IRAS observations which are
available for all but seven of our sources \citep{Neugebauer1984}. 

 Using interstellar extinction estimates from either the model by
\cite{Arenou1992} or from the observed extinction towards the
early-type stars in the corresponding clusters, we deredden the
photometric and spectroscopic data with the law presented by
\cite{Cardelli1989}, extended toward longer wavelengths with the local
ISM curve of \cite{Chiar2006}, and assuming R$_{\rm{V}}=3.6$
\citep{Massey2005}.

\section{Modelling}

The focus of this study is to derive acurate dust masses and its
composition. Therefore, several aspects of the complex RSG atmospheres
are not taken into account and the following simplifications are made
constructing the radiative transfer models:
(1) all calculations are done in 1D-spherical geometry, which means we neglect
inhomogeneities, (2) we do not take the chromosphere into account,
(3) we do not include any extra-photospheric molecular layers and (4)
we assume the mass loss to be constant.

\subsection{Photospheres}

The stellar photospheres are represented by {\sc marcs} (edition 1998) models
\citep{Gustafsson2008} as these are specifically
developed for cool stars, with an emphasis on molecular
opacities and the effects of extended atmospheres which do not allow a
plane-parallel approximation. They were computed in spherical
geometry with solar abundances, a surface gravity\footnote{Lower gravity models failed to converge at the lowest
temperatures which led us to adopt this value uniformly. The stellar
mass of the models in this grid is only 1\,M$_{\odot}$, but at the
resolution of our set of observation, the difference with higher-mass
models would not be noticeable.} of $\log{g} =
0.5$, a
microturbulent velocity v$_{\rm{turb}}=2$\,km\,s$^{-1}$ and
temperatures ranging from 3000 to 4500\,K. The effective temperatures
were derived from the spectral type following the relations from
\cite{Levesque2005}.

\subsection{Dust radiative transfer}

The dust shell is modelled using the proprietary spherical radiative
transfer code {\sc modust} \citep{Bouwman2000,Bouwman2001}. Under the
constraint of radiative equilibrium, this code solves the
monochromatic radiative transfer equation from UV/optical to
millimetre wavelengths using a Feautrier type solution method
\citep{Feautrier1964, Mihalas1978}. The code allows to have several
different dust components of various grain sizes and shapes, each with
its own temperature distribution.

\subsection{Dust composition}

We searched the literature for relevant dust species to include in our
study. \cite{Speck2000} find in the RSG in their sample melilite,
alumina and olivines\footnote{Remark that the term Olivine in
principle refers to the crystalline form. Although the correct
phrasing should be ``amorphous silicate with an olivine
stoichiometry'', we refer to this material as Olivines, as is done in
similar studies.} , but they do not need a 13\,$\mu$m-carrier
candidate such as Spinel as they do not detect this feature in their
RSG spectra and neither do we (see Sect.\,\ref{sec:residuals}).
\cite{Heras2005} find in their sample of AGB stars those species used
by \cite{Speck2000} and also Mg$_{0.1}$Fe$_{0.9}$O, which has a
feature at 19.5\,$\mu$m, outside the wavelength band observed by
\cite{Speck2000}. \cite{Cami2002} uses olivines, alumina, spinel and
Mg$_{0.1}$Fe$_{0.9}$O. For a discussion as to the need for melilite
instead of an iron-magnesium silicate in RSG, we refer to
\cite{Speck2000}, who show the existence of a category of RSG
10\,micron spectra with a peak position significantly to the red of
that of olivines. Moreover, they show an excellent agreement with
a predicted melilite emission spectrum. 

 From a theoretical point of view, a dust formation scenario
  usually starts with TiO$_2$ seeds, on which first simple
  oxides such as Al$_2$O$_3$ and MgFeO grow. If the
  density/temperature conditions are right, also melilite, and
  olivines and pyroxenes with iron content can condense onto the
  grains. The full condensation scheme consists of quite a few more
  possible dust species (e.g. Anorthite and Plagioclase), but no
  spectral sign of these dust types has so far been seen in RSG.

The dust species used in this study and their relevant
characteristics are presented in Table\,\ref{tab:dust}. The optical
constants are obtained from the AIU Jena
database\footnote{http:$//$www.astro.uni-jena.de$/$Laboratory$/$Database$/$databases.html}.

\begin{table*}
 \begin{center}
 \caption{Dust species used in this study with their commonly-used
 name, their size, shape and the source of the optical constants. The
 size distribution is that derived by \cite{Mathis1977} and the cross
 section are computed in the CDE (Continuous Distribution of
 Ellipsoids) approximation. 
   \label{tab:dust}
  }
 \vspace{1ex}
 \begin{tabular}{lccccc}
  \hline
  \hline
  Name           & Composition  & Lattice  structure  & Size (N($\alpha$)$\sim\alpha^{-3.5}$) & Shape &
  Reference \\
\hline
  Alumina        &  Al$_2$O$_3$  & amorphous      & $0.01-1 \mu$m & CDE &
  \cite{Begemann1997,Koike1995} \\
  Melilite       &  Ca$_2$Al$_2$SiO$_7$ & amorphous & $0.01-1 \mu$m & CDE &
  \cite{Mutschke1998,Jaeger1994} \\
  Olivine        &  Mg$_{0.8}$Fe$_{1.2}$SiO$_4$ & amorphous & $0.01-1 \mu$m & CDE &
  \cite{Dorschner1995} \\
  MgFeO          &  Mg$_{0.1}$Fe$_{0.9}$O & amorphous & $0.01-1 \mu$m & CDE &
  \cite{Henning1995} \\
  Metallic iron  &  Fe                & crystalline  & $0.01-1 \mu$m & CDE &
  \cite{Henning1996} \\
  Carbon         &  C                 & amorphous & $0.01-1 \mu$m & CDE &
  \cite{Preibisch1993} \\
 \hline
  \end{tabular}
 \end{center}
\end{table*}

The RSG in this sample appear to be pure outflow sources, i.e. nothing
indicates that they may be binary systems with stable circumbinary
dust disks: as shown in Fig.\,\ref{fig:fits}, there is no strong
discrepancy between reddening and IR excess, nor do we see any sign of
processed, crystalline grains such as those seen around some Post-AGB
objects \citep[e.g. ][]{Gielen2007} or in protoplanetary disks
\citep[e.g. ][]{vanBoekel2005}. This implies that grain growth beyond
0.1\,$\mu$m is unlikely \citep[see e.g. ][]{Gail1999,Woitke2005}. The
absorption cross sections can therefore be calculated in the Rayleigh
limit, and an underlying grain size distribution is only required to
derive the dust masses. We use the size distribution derived by
\cite{Mathis1977} and compute the cross section in the approximation
of a Continuous Distribution of Ellipsoids \citep[CDE,][]{Bohren1983}, as we found this to
provide a better agreement with the observations than the Mie
approximation. More sophisticated treatments of grain sizes and shapes
exist now, such as a Distribution of Hollow Spheres (DHS), but within
the Rayleigh limit, CDE works very well \citep{Min2005}. We must
point out that according to very recent work by \cite{Hoefner2008}, relaxing the
restriction to the small particle limit in the modelling of AGB winds leads to a possible grain
growth up to 1\,$\mu$m, which may also help in the radiative driving
of the wind. 

\begin{figure}
\centering
  \resizebox{\hsize}{!}{\includegraphics{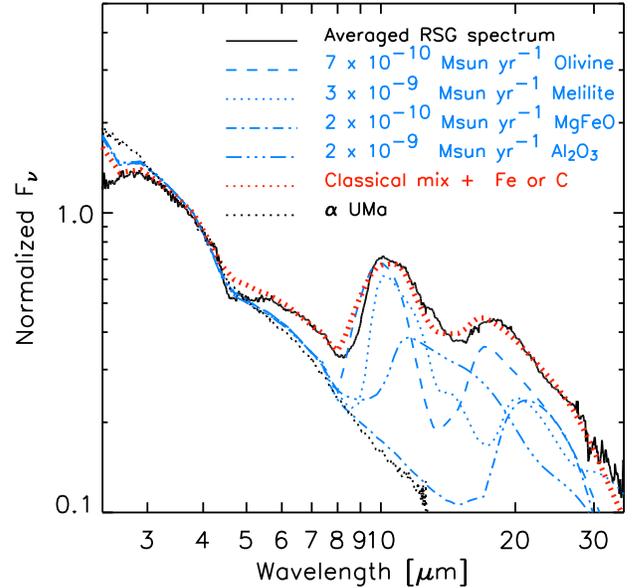}}
  \caption{The average of all dereddened observed dust spectra, each
    normalized with its photospheric flux at 3.5\,$\mu$m. For
    comparison, the pure photospheric spectrum of $\alpha$\,UMa is
    shown as a black dotted line. The spectral features of the
    commonly used dust species are also shown (blue) in the way they
    appear in models with CSEs containing only that dust type. None of
    these dust species can explain the excess already observed before
    8\,$\mu$m. The red line is a model containing also some Fe and/or
    C. 
}
  \label{fig:features}
\end{figure}

An overview of
the spectral features of these different dust species is shown in
Fig.\,\ref{fig:features}.  An in-depth discussion of
this figure is presented in Sect.\,\ref{sec:sum}. 

%
%
%

\subsection{The strategy}

The unknown angular scale of the central star is determined by scaling
the model to the ISO-SWS spectrum between 3.5 and 4.5\,$\mu$m as these
wavelengths are relatively free of dust and molecular
emission/absorption.  This method allows us to bypass the uncertain
absolute luminosity and distance. The geometrical properties of the
dust shell are proportional to the stellar radius and thus their
angular scale is now also fixed. Under the assumption of a continuous
wind, the outer radius of the dust shell
can not be determined from the current modelling as the outer regions
of the wind are too cold to contribute to the Mid-IR spectrum. We
therefore fix its value at 5000\,R$_{\star}$. 

The mass-loss rate does depend on the assumed distance and
luminosity. The initial model grid (central star + dust shell) is calculated for a typical
luminosity of 10$^5$\,L$_{\odot}$. This model mass-loss rate
$\dot{\rm{M}}_{\rm{grid}}$ is
converted to an actual $\dot{\rm{M}}$ using the distances presented in
Sect.~\ref{sec:sample} and the angular scale derived from the fitting
(see also Sect.\,\ref{sec:results}). The outflow velocity
and dust-to-gas ratio are assumed to be 10\,km\,s$^{-1}$ and 0.01
respectively, both typical values for RSG. Note that these are
uncertain to a factor of at least 2.

The parameters to be determined from the individual fitting of the
spectra are the inner radius, R$_{\rm{in}}$, the mass-loss rate,
$\dot{\rm{M}}$, and the relative abundances of the different dust species. 

We refrain from doing a formal minimisation while determining the
optimum dust-shell parameters. Instead we determine the best-fit model
using visual inspection, following a well defined strategy. The
reasons for doing so are two-fold. {\it 1)} It is our aim to get, at
the same time, a good fit to the SED {\bf and} the detailed
IR-spectrum. This is a requirement as the circumstellar reddening
puts a strong constraint on the inner shell radius, which in turn
determines the relative contribution of the individual dust species to
the observed features through their different temperatures. It is
virtually impossible to devise a good weighting scheme to give equal
importance to the $\sim$10 photometric points and the thousands of
spectral points in the SWS spectra. In particular, since the
uncertainties on the data vary considerably throughout the sample,
this would imply adapting the weighting on a per source basis.  The
only way to judge whether such a complicated weighting scheme works
well would be by visual inspection. {\it 2)} There are some spectral
regions that may be poorly reproduced even with a, to our judgement,
good model. This is due to the possible presence of components in the
real spectra which are lacking in the model, like extended molecular
layers (also known as MOLspheres) or
PAHs. Again, such regions and there importance can only be identified
by visual inspection.

The core of our strategy is an iteration of the following
sequence:
\begin{itemize}
\item We estimate the dust mass-loss rate from the full SED and the
  observed
  L$_{\rm{IR}}$/L$_{\star}$\footnote{{\bf L$_{\rm{IR}}$ is calculated as
  the integration over $\lambda \ge 2\,\mu$m of the final model
  spectrum minus the photospheric model and L$_{\star}$ is calculated
  by integrating the photospheric model over all wavelengths.}}.
\item The relative abundances of melilite, olivine and alumina  are
  adjusted to match the peak position of the 10\,$\mu$m feature
\item  and the same is done for the 18\,$\mu$m feature peak position
  using olivine, melilite and Mg$_{0.1}$Fe$_{0.9}$O.
\item When the peak positions of the features are well reproduced but
  not yet their strength, we vary the inner radius to adjust their
  relative strenghts and the mass-loss rate to alter the absolute
  emission levels. Meanwhile, we check that our model predicts the
  correct amount of circumstellar reddening.  
\end{itemize}

\section{Results}

\subsection{A continuum opacity source}
\label{sec:sum}

In Fig.\,\ref{fig:features}, we present an ``averaged'' RSG ISO-SWS
spectrum, where each individual spectrum is dereddened with the
appropriate value (last column in Table\,\ref{tab:sample}) and
normalized to the average flux between 3.5 and 4\,$\mu$m. We exclude
RSG  without significant mass loss. Also shown are models with
envelopes containing only a single dust type. 

%
%
%
%
%

Striking in this figure is that the observed excess appears to start
already at 5\,$\mu$m, while the modelled dust features are all located
beyond 8\,$\mu$m. A related problem could be the slope of the
photospheric(?) part of the modelled spectrum at 2.5--3\,$\mu$m. 

The excess around 6\,$\mu$m and the difference in slope around
3\,$\mu$m have recently been attributed to the presence of a
MOLsphere, i.e. an extra-photospheric layer of molecular material
\citep[e.g. ][]{Tsuji2000}, in which especially H$_2$O would be the
main source of excess emission and absorption. Individual lines of
water in these MOLspheres have been detected
\citep[e.g. ][]{Jennings1998} and the shells have been resolved by
near and mid-IR interferometry \citep[e.g.
][]{Perrin2004,Ohnaka2004,Perrin2005}. \cite{Verhoelst2006} argue that
a purely molecular MOLsphere cannot explain the wavelength dependence
of the opacity across the entire IR range in the case of RSG
$\alpha$\,Ori and they suggest alumina as additional mid-IR opacity
source. However, this does not yet explain the observation made by
\cite{Ryde2006} in the spectra of $\mu$\,Cep, that water lines appear
in absorption at 12\,$\mu$m, while they are assumed to generate an
excess at 6\,$\mu$m\footnote{The MOLspheres are found to be colder than
the photospheric background, and generate excess emission only through their
larger emitting surface.}. Moreover, the excess around 6\,$\mu$m in
the individual stars does not show any water lines in proportion to
the excess w.r.t. the photospheric continuum.

We propose here that, although extra-photospheric molecular material
is clearly present, the peculiarities in the shape of the overall SED
discussed here are due to a source of continuous opacity with a fairly
cool temperature. Free-free
emission from a chromosphere or ionised wind can be present in
RSG \citep[e.g.][]{Harper2001}, but since its source function has a
temperature above that of the photosphere, it can not explain the
additional extinction towards the near-IR\footnote{We know that the
  near-IR is suffering additional extinction (as oppposed to less
  excess emission) from the fit of our photosphere models to the
  unreddened optical fluxes}.  The color temperature of the excess
therefor points to dust rather than free-free emission. We
find that the inclusion of either metallic Fe, amorphous C or
micron-sized grains in the
dust shell can explain the excess emission at 6\,$\mu$m, the slope
issue at 3\,$\mu$m and the additional extinction at near-IR
wavelengths (see Sect.\,\ref{sec:results}). The remaining residuals are
molecular absorption, as expected from the presence of a cool
MOLsphere.

Whether this missing dust species is either metallic iron,
micron-sized grains or amorphous
carbon, can not be fully answered with the current analysis. The
presence of metallic Fe in the CSE of VY\,CMa has been postulated by
\cite{Harwit2001} and it is also used to explain the SEDs of OH/IR
stars \citep{Kemper2002}. Moreover, metallic Fe is a product of the
O-rich condensation sequence.  Nevertheless, amorphous carbon is an
equally interesting possibility, as it would be roughly 20 times as
effective in accelerating the wind as Fe \citep[see Table\,1 in
][]{Woitke2006}. The occurence of pure carbon in an oxygen-rich CSE is
plausible \citep{Hoefner2007}, even more so in RSG where the
chromospheric radiation field can be responsible for the dissociation
of CO. We find that, for the amount of pure carbon required to explain
the RSG spectra, only of the order of 1\,\% of the CO must be
dissociated. This is well below what is found possible in theoretical
models of M stars with chromospheres \citep{Beck1992}. More evidence
for the presence of carbon not bound in CO is the detection of PAH
emission in four of our sources (Sect.\,\ref{sec:PAHs}).  Papers
discussing the possible presence of amorphous carbon in oxygen-rich
outflows often dismiss the spectroscopic detection of its presence as
impossible due to the lack of spectral features. Interestingly, we
find here that, when studying the full SED, it is possible to detect
continuous opacity sources. Continuous opacity by micron-sized O-rich grains, playing a crucial role
in the driving of the wind, has been proposed for AGB stars by
\cite{Hoefner2008}, but it is unclear whether this grain growth can
also occur in RSG.

Regardless of the origin of this continuum emission, it must be
included in our modelling to derive reliable mass fractions for the
other dust species. We have chosen to include amorphous C, in the
knowledge that Fe would have a similar effect.

\subsection{Models for all sample stars}
\label{sec:results}

\begin{figure*}
\centering
  \resizebox{\hsize}{!}{\includegraphics{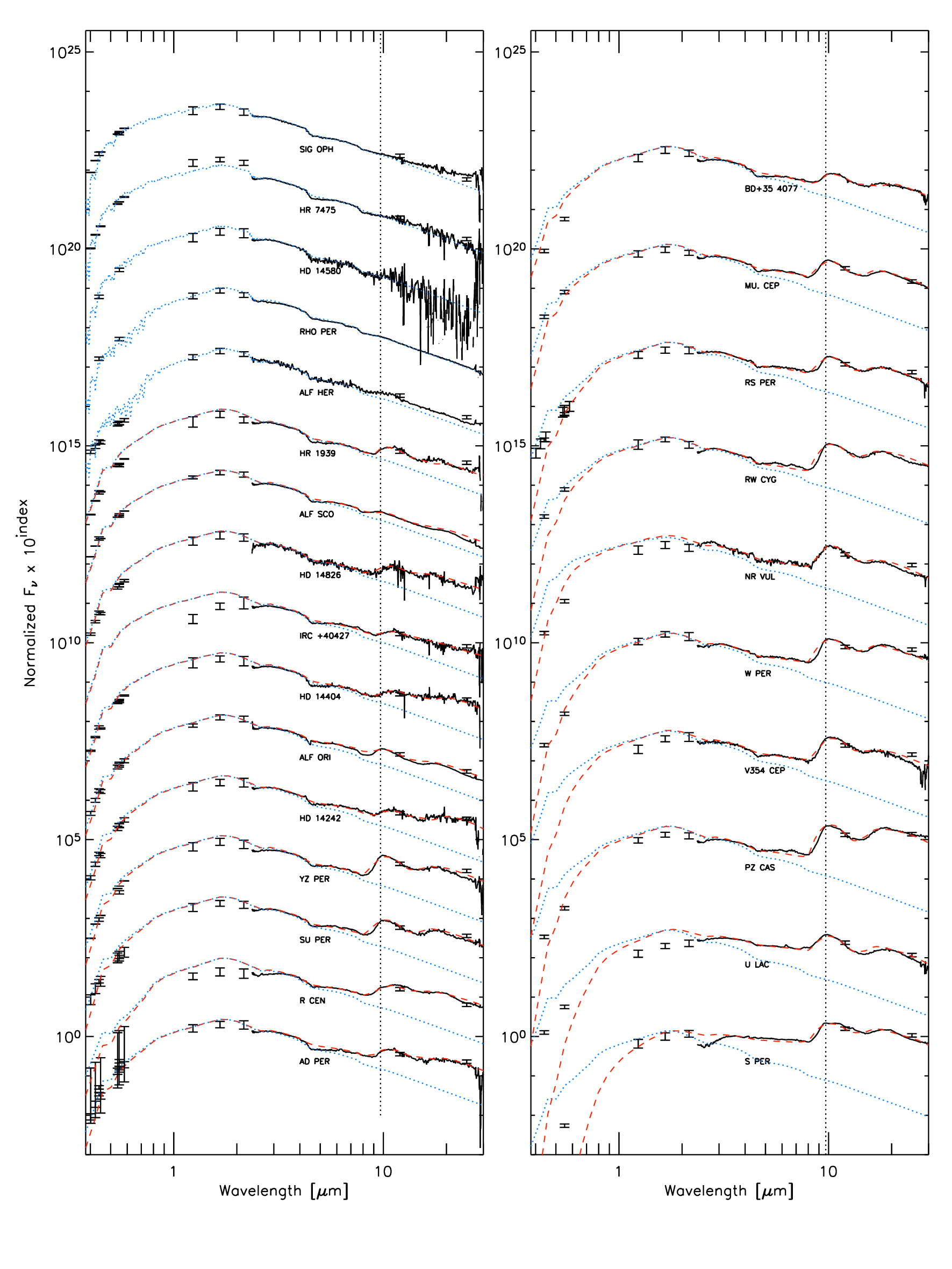}}
  \caption{The SEDs of all sample stars with their best-fit models
    (total spectrum in red dashes and non-reddened photospheric
    spectrum in bluedots) The photometric points are either the 7
    Geneva filters or B and V in the Johnson system, J H and Ks from
    2mass and I12 and I25 from the IRAS mission. The dotted vertical
    line indicates 9.7\,$\mu$m, the wavelength where ``classical''
    amorphous silicates exhibit their peak.
}
  \label{fig:fits}
\end{figure*}

\begin{figure*}
\centering
  \resizebox{\hsize}{!}{\includegraphics{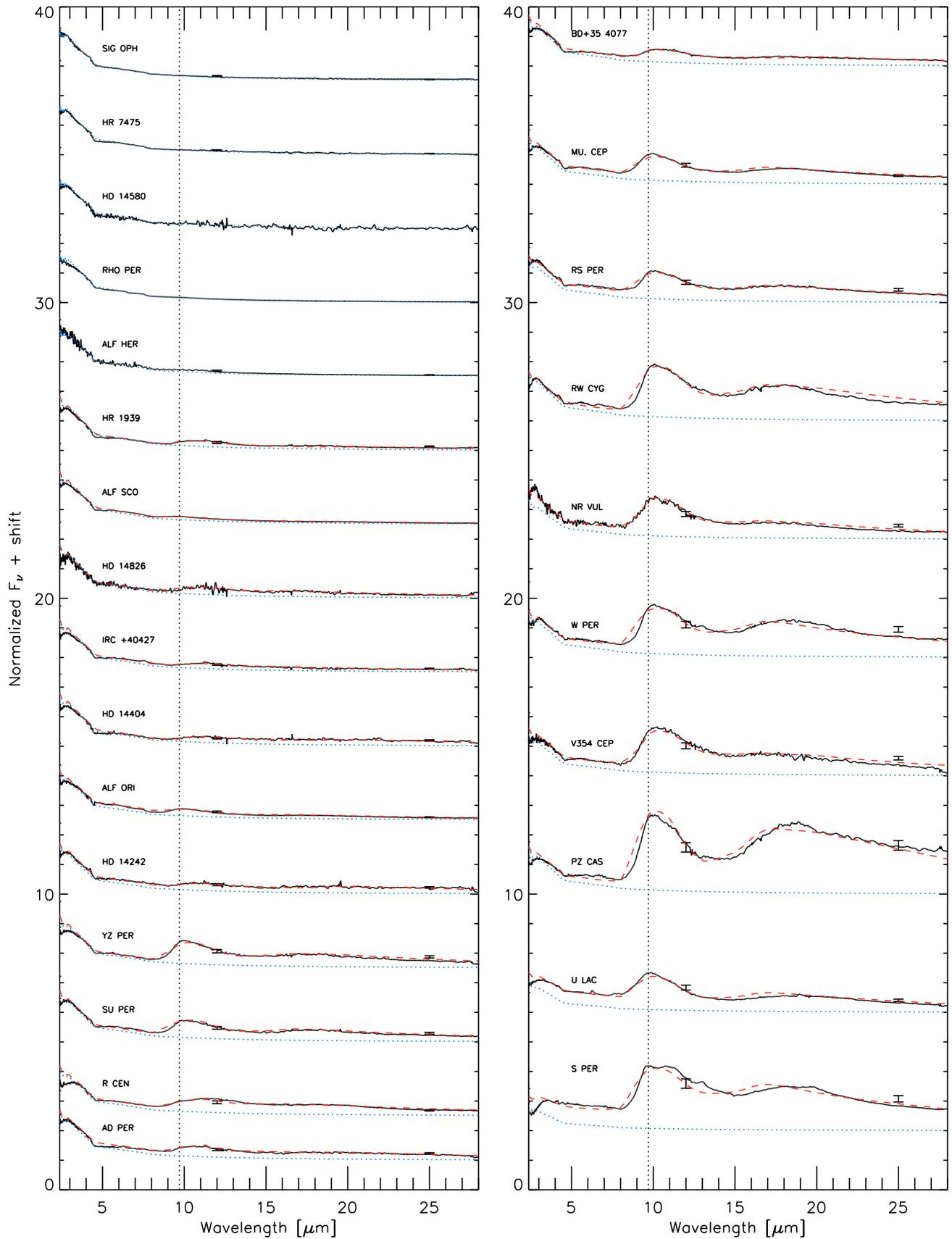}}
  \caption{The ISO-SWS spectra of all sample stars with their best-fit
  models (symbols and colours as in Fig.\,\ref{fig:fits}).
    }
  \label{fig:isofits}
\end{figure*}

Figures\,\ref{fig:fits} and \ref{fig:isofits} present the observed
SEDs and spectra respectively and the best-fit
models, sorted following increasing dust luminosity or following
decreasing photospheric temperature in case no dust emission is
present. In general the agreement is excellent.  IRC\,+40427 appears
much more reddened than our estimate of $A_V = 1.79$, which suggests
it to be located much further away than our default value of 2\,kpc.
%
%
%
%
%
%
%

The results on all 27 sample stars are summarized in
Table\,\ref{tab:results}. For 6 objects, the mass-loss rate is too low
($\dot{\rm{M}} \le 10^{-9} \rm{M}_{\odot}$\,yr$^{-1}$) to allow for a
reliable determination of the dust composition.
The distance and
luminosity corrected mass-loss rate presented in column\,9 is computed as
$$\dot{M}=\dot{M}_{\rm{grid}}\,A\, (d/1\,\rm{kpc})^2$$ where
$\dot{M}_{\rm{grid}}$ is the mass-loss rate for an assumed luminosity
of 10$^5$\,L$_{\odot}$ (the default value in our model grid), $A$ is the scaling factor used to match the
observed flux around 4\,$\mu$m and $d$ is the distance.

\begin{table*}
 \begin{center}
 \caption{
The best fit model parameters.
   \label{tab:results}
  }
 \vspace{1ex}
 \begin{tabular}{llcccccccccc}
  \hline
  \hline
  Source & Spectral & T$_{\rm{eff}}$  & Melilite & Olivine  & Alumina &  Mg$_{0.1}$Fe$_{0.9}$O & Carbon & R$_{\rm{in}}$    & $\dot{\rm{M}}_{\rm{model}}$       & $\dot{\rm{M}}$         & L$_{\rm{IR}}$/L$_{\star}$ \\
         & Type     &     [K]         & \%-mass  &  \%-mass  & \%-mass & \%-mass & \%-mass &   [R$_{\star}$] &  [M$_{\odot}$ yr$^{-1}$] &  [M$_{\odot}$ yr$^{-1}$]  &  \\
  \hline
     NR VUL &       K3Iab & 4015 &     0.56 &     0.19 &     0.24 &     0.00 &    0.013 & 10 &    3.1e-07 &    2.7e-07 &    0.074   \\
{\sl    SIG OPH}   &   K3Iab   &4015   &   &   &   &   &   &   &   &   &  \\
    HR 7475   &    K4Ib   &3900   &   &   &   &   &   &   &   &   &  \\
 IRC +40427 &      M1:Iab & 3745 &     0.43 &     0.00 &     0.55 &     0.00 &    0.018 & 15 &    1.1e-07 &    5.0e-08 &    0.011   \\
   HD 14580   &   M1Iab   &3745   &   &   &   &   &   &   &   &   &  \\
   HD 14404 &       M1Iab & 3745 &     0.48 &     0.00 &     0.48 &     0.04 &    0.007 & 15 &    1.5e-07 &    7.7e-08 &    0.012   \\
    ALF SCO &    M1.5Iabb & 3710 &     0.74 &     0.09 &     0.12 &     0.00 &    0.049 & 13 &    4.1e-08 &    3.3e-08 &    0.009   \\
    MU. CEP &        M2Ia & 3660 &     0.31 &     0.27 &     0.41 &     0.00 &    0.020 & 18 &    2.9e-07 &    7.9e-07 &    0.048   \\
{\sl   HD 90586} &    M2Iab/Ib & 3660 &     0.57 &     0.14 &     0.23 &     0.05 &    0.011 & 10 &    1.4e-07 &    8.5e-09 &    0.020   \\
    ALF ORI &       M2Iab & 3660 &     0.64 &     0.16 &     0.20 &     0.00 &    0.001 & 13 &    6.3e-08 &    2.9e-08 &    0.022   \\
{\sl    HR 1939} &       M2Iab & 3660 &     0.82 &     0.01 &     0.16 &     0.00 &    0.003 & 10 &    1.1e-07 &    6.1e-09 &    0.006   \\
     YZ PER &       M2Iab & 3660 &     0.60 &     0.14 &     0.26 &     0.00 &    0.003 & 18 &    3.5e-07 &    2.7e-07 &    0.026   \\
   HD 14826 &       M2Iab & 3660 &     0.42 &     0.05 &     0.52 &     0.00 &    0.005 & 15 &    1.7e-07 &    1.0e-07 &    0.010   \\
   HD 14242 &       M2Iab & 3660 &     0.59 &     0.00 &     0.34 &     0.05 &    0.017 & 15 &    1.8e-07 &    7.2e-08 &    0.023   \\
   V354 CEP &     M2.5Iab & 3615 &     0.57 &     0.14 &     0.28 &     0.00 &    0.008 & 11 &    5.6e-07 &    9.8e-07 &    0.087   \\
 BD+35 4077 &     M2.5Iab & 3615 &     0.78 &     0.03 &     0.16 &     0.02 &    0.010 & 15 &    2.9e-07 &    2.0e-07 &    0.029   \\
     AD PER &     M2.5Iab & 3615 &     0.59 &     0.03 &     0.34 &     0.02 &    0.017 & 15 &    2.4e-07 &    1.5e-07 &    0.029   \\
     RW CYG &       M3Iab & 3605 &     0.53 &     0.17 &     0.30 &     0.00 &    0.002 & 25 &    1.0e-06 &    9.4e-07 &    0.069   \\
     PZ CAS &       M3Iab & 3605 &     0.40 &     0.30 &     0.30 &     0.01 &    0.002 & 40 &    1.7e-06 &    5.4e-06 &    0.109   \\
     SU PER &       M3Iab & 3605 &     0.39 &     0.19 &     0.39 &     0.03 &    0.008 & 15 &    2.6e-07 &    3.0e-07 &    0.027   \\
      U LAC &     M4Iab:e & 3535 &     0.25 &     0.44 &     0.25 &     0.01 &    0.053 & 10 &    3.7e-07 &    9.0e-07 &    0.154   \\
{\sl    RHO PER}   &    M4II   &3535   &   &   &   &   &   &   &   &   &  \\
      W PER &     M4.5Iab & 3535 &     0.09 &     0.27 &     0.63 &     0.00 &    0.009 & 20 &    7.8e-07 &    6.4e-07 &    0.085   \\
      S PER &     M4.5Iab & 3535 &     0.03 &     0.27 &     0.68 &     0.01 &    0.022 & 15 &    1.8e-06 &    2.2e-06 &    0.302   \\
     RS PER &       M4Iab & 3535 &     0.42 &     0.21 &     0.31 &     0.04 &    0.017 & 13 &    3.8e-07 &    4.2e-07 &    0.063   \\
    ALF HER   &   M5Iab   &3450   &   &   &   &   &   &   &   &   &  \\
      R CEN &    M5IIevar & 3450 &     0.33 &     0.06 &     0.60 &     0.00 &    0.009 & 15 &    3.7e-07 &    3.3e-07 &    0.028   \\
\hline
 \end{tabular}
 \end{center}
\end{table*}

\section{Analysis of the residuals}
\label{sec:residuals}

A general problem, especially for the higher mass-loss stars, is
the over-prediction of the near-IR photometry. This corresponds to the
need for an extinction law with a higher $R_V$. The dust already
includes a component with grey extinction properties (see
Sect.\,\ref{sec:sum}), but in our models it is located too far from
the central star to provide sufficient extinction. Our model assumes
the same inner radius for all dust species, but the discrepancy
observed here may indicate that the continuous opacity source is
located even closer to the base of the wind, right above the
photosphere. The detection of alumina 0.5\,R$_{\star}$ above the
photosphere of Betelgeuse
\citep{Verhoelst2006,Perrin2007} shows that dust condensation so close
to the photosphere is possible.

\subsection{Molecular bands}
\label{sec:mol}

\begin{figure}
\centering
  \resizebox{\hsize}{!}{\includegraphics{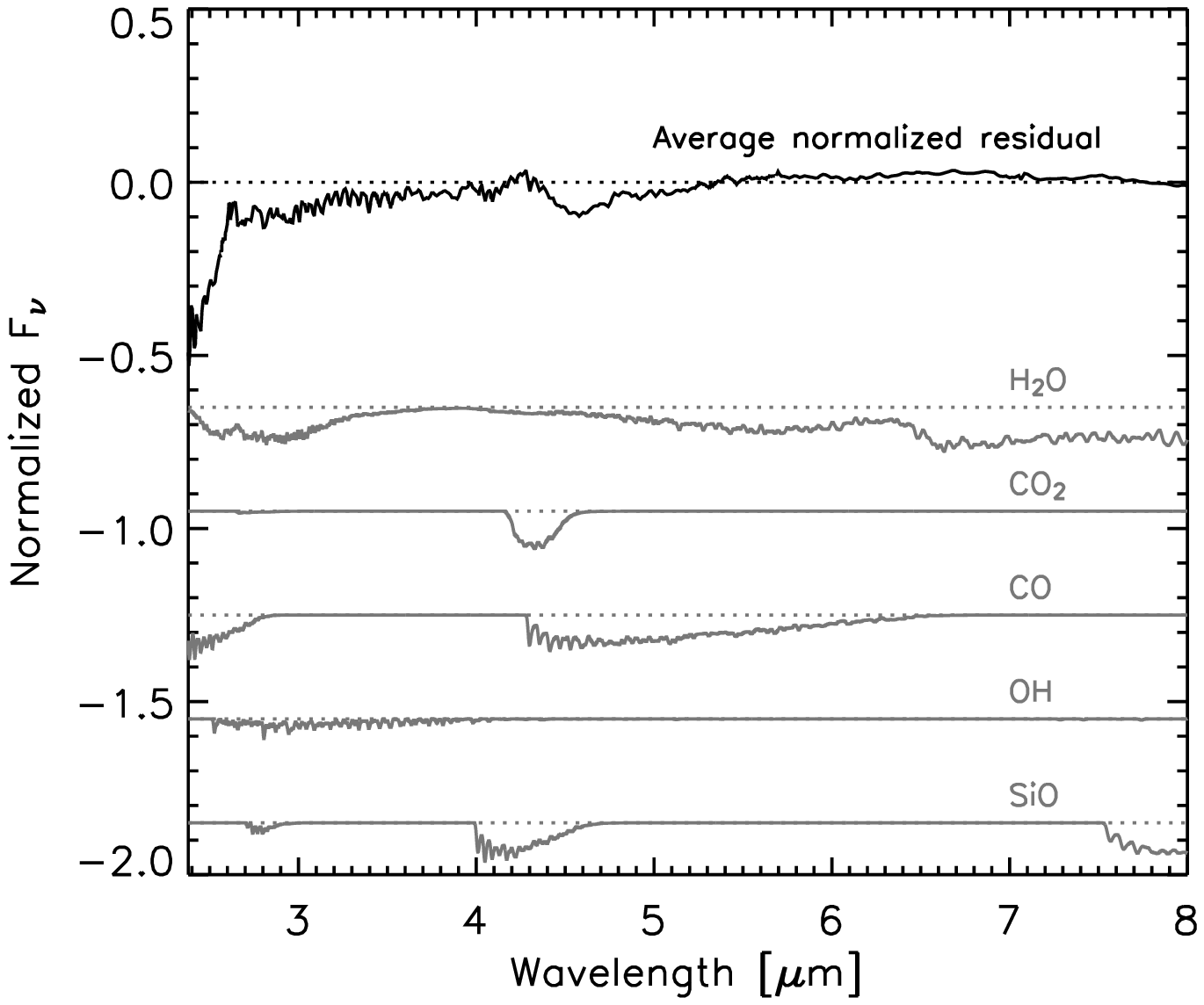}}
  \resizebox{\hsize}{!}{\includegraphics{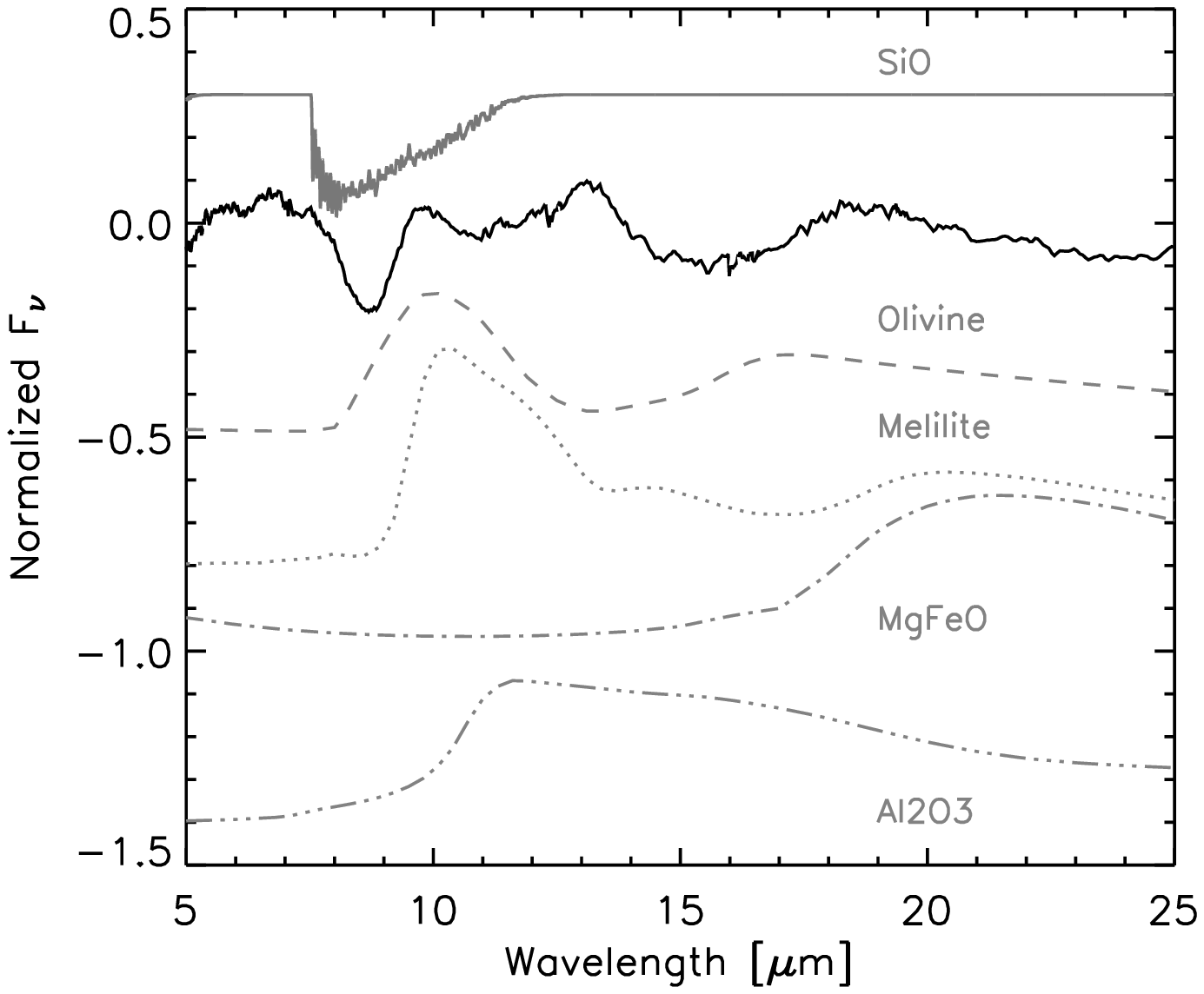}}
  \caption{{\sl Upper panel:} The average normalized residuals in the blue part of the
    ISO-SWS spectrum together with simulated molecular absorption bands.
    For more details on the computation of the molecular templates we
    refer to \cite{Cami2002}. {\sl Lower panel:} The wavelength range of the
    residuals (black solid line) which is dust dominated. }
  \label{fig:residuals}
\end{figure}

The ISO spectra up to 8\,$\mu$m are dominated by photospheric and
possibly continuum dust emission. The observed spectral features
should therefore mainly be photospheric molecular absorption bands. To
search for evidence for the presence of the MOLsphere(s) discussed in
Sect.\,\ref{sec:sum} or for indications of shortcomings in our model
atmospheres, we compare the average normalized\footnote{Normalization
  of each observed and model spectrum was done by inverse scaling with
the average flux (F$_{\nu}$) between 3.5 and 4\,$\mu$m. The spectra were then
subtracted, and the residuals calculated in this way averaged over the
sample.} residuals with
simulated molecular absorption bands in Fig.\,\ref{fig:residuals}.  
Although our models clearly match the observed spectra very well (the
better part of the residuals does not reach a 5\,\% level), we detect
strong additional absorption by CO and maybe also some water and
OH. The latter two discrepancies can probably be solved by more
detailed fine tuning of the atmosphere models. Cool CO appears to be very
abundant also above the photosphere, which is in line with a
postulated crucial role in the wind-driving mechanism. The fact that
no massive MOLspheres containing water are observed clearly sets the
RSG apart from the AGB stars. This finding supports our assumption
(Sect.\,\ref{sec:sum}) that the excess at 6\,$\mu$m and the slope
issue at 3.5\,$\mu$m is due to some continuum emission source and not an
extra-photospheric water column.  
%
%
%

\subsection{Residual dust features}

The lower panel of Fig.\,\ref{fig:residuals} shows the average
residuals in the dust-dominated part of the spectrum. Many
discrepancies are clear: (1) a strong absorption band starting at
8\,$\mu$m and possibly extending up to 12\,$\mu$m if we interpret the
emission bump at 9.7\,$\mu$m as an underestimated olivine dust
fraction, (2) a broad 13\,$\mu$m feature and (3) a broad emission
feature at 18\,$\mu$m. 
Concerning the first problem: a significant SiO layer just above the photosphere, which deepens the
absorption band, could be present as such a layer is detected around $\alpha$\,Ori with
 MIDI interferometry by \cite{Perrin2007}, but the SiO bandhead does
 not correspond very well with the observed residuals around
 8\,$\mu$m. It may also be that the current dust model does not yet 
 contain the right silicate dust composition/shape/grain size.  

\subsubsection{A 13 $\mu$m feature?}

\cite{Speck2000} find in their analysis that RSG usually do not
display the 13\,$\mu$m feature as seen in AGB stars. We find in
general, after model substraction,  a feature
around 13\,$\mu$m, but it is much broader than the one seen in the
lower-mass counterparts. Only a few selected targets display the
AGB-like 13\,$\mu$m feature, but these stars (SU Per and S Per) may be
different in other aspects as well: S Per is the only target in our
sample showing strong water bands, and \cite{Speck2000} already
suggested it to be more similar to lower-mass semi-regulars. The broad
excess in the residual spectrum can not be isolated in any of the
spectra prior to model subtraction. Therefore, it can not be
positively identified as a component, but might be due to some
systematic problem with our models. The most likely systematic effect
to cause such an underprediction would be a lack of absorptivity in
one of the sets of optical properties that we have used.

\subsubsection{The 18 $\mu$m feature}

The 18\,$\mu$m feature observed in the residuals is most probably the
feature typical ascribed to the silicates generating the 9.7\,$\mu$m
feature (the olivines). However, starting from laboratory-measured
optical constants, we find it impossible to reproduce this peak
position: either it is blue shifted (olivines) or it peaks to the red
(melilite). The inclusion of either alumina or MgFeO causes the model
emission band to be centered at the observed wavelength. However, this
always causes a double peaked band profile which is never observed. We
note that this discrepancy is not unique to the RSG but is also found
in the AGB stars with optically thin dust shells as studied by
\cite{Heras2005}. This problem may be related to that at 8\,$\mu$m.

\subsubsection{Detection of PAHs}
\label{sec:PAHs}
\begin{figure}
\centering
  \resizebox{\hsize}{!}{\includegraphics{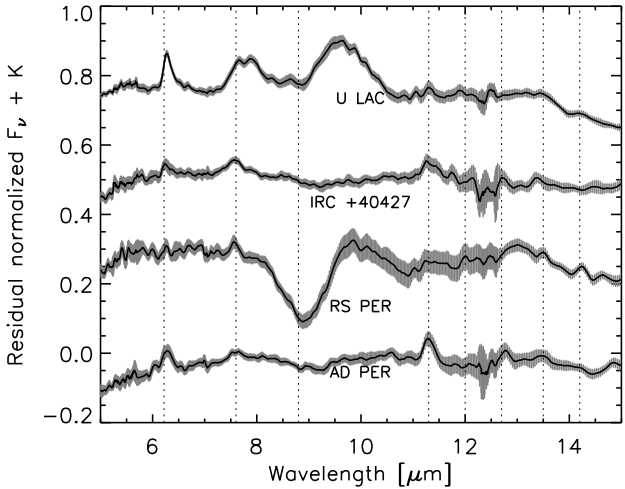}}
  \caption{The 5 to 15 $\mu$m spectra of RSG showing PAH
  emission. The grey error bars correspond to 3\,$\sigma$. Known PAH feature wavelengths are indicated with a dashed line. These wavelengths
  are taken from \cite{Hony2001} and \cite{Peeters2002}.}
  \label{fig:PAHs}
\end{figure}

As shown in Fig.\,\ref{fig:PAHs}, we clearly detect PAH emission
bands at 6.3, 7.6, 11.3 and 14.2\,$\mu$m in 3 stars: AD\,Per,
IRC\,40427, and U\,Lac. RS\,Per only shows a significant PAH feature
at 7.6\,$\mu$m. AD\,Per and IRC\,40427 also show a feature at
12.7\,$\mu$m.  The PAHs in AD\,Per, RS\,Per and IRC\,40427 were
already detected by \cite{Sylvester1994,Sylvester1998} who relate
their presence with the dissociation of CO molecules by the UV
radiation field of the chromosphere, as no PAHs are detected in AGB
stars. The diffuse ISM is known to show PAH emission features but we
are confident that the bands observed here originate in the CSE of the
RSG since 1) the observed bands are as strong in the UKIRT
observations of \cite{Sylvester1994} as in our ISO-SWS observations,
in spite of a different aperture, 2) the UKIRT observations used a
chopping technique, which should remove a large fraction of an
extended ISM feature and 3) the relative band strenghts do not agree
with those of the ISM and the 6.22\,$\mu$m feature is, in some cases,
shifted to longer wavelengths.

Using the
dust shell models presented here, we are able to isolate the PAH
emission features. This allows us to study the band shapes and
strengths in a manner similar to \cite{Peeters2002} and
\cite{Hony2001}. Such an analysis can yield important information on
the structural properties of the PAH family and give clues about their
formation conditions. This analysis is left to a future study. Perusal
of Fig.\,\ref{fig:PAHs} already shows that the PAH features in our
sample span an interesting range both in which bands are observed,
band strength ratios and band-shapes. 
%
%
%

%
%
%
\begin{figure*}
\centering
  \resizebox{\hsize}{!}{\includegraphics{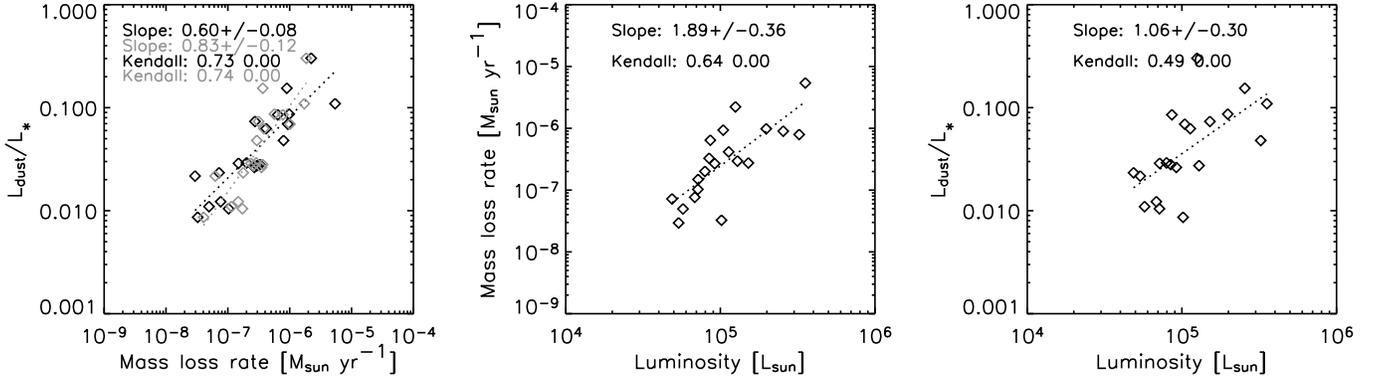}}
  \caption{{\it Left panel: }The correlation between mass-loss rate and the ratio of dust to
    photospheric luminosity. The slope is that of a linear fit, the
    Kendall values refer to the Kendall rank correlation value and its
    significance. The Kendall rank correlation $\tau \equiv \frac{n_c
      - n_d}{\frac{1}{2}n(n-1)}$ where $n$ is the size of the sample, $n_c$ is the number of
    concordant pairs and $n_d$ is the number of discordant pairs. The
    significance is a value in the interval [0.0, 1.0] where a small value indicates a significant correlation. The relation is clear using both
    $\dot{\rm{M}}_{\rm{grid}}$ (grey symbols) and the
    distance-corrected mass-loss rate (black symbols). {\it Middle
    panel: } Mass-loss rate (corrected for luminosity and distance)
    versus stellar luminosity. {\it Right panel: } The luminosity- and
    distance-independent mass-loss indicator versus stellar
    luminosity. Only stars which are certainly RSG were used to
    measure these correlations.} 
  \label{fig:spcorrelations}
\end{figure*}  

\section{Analysis}
\subsection{L$_{\rm{dust}}$/L$_{\star}$ as mass-loss rate indicator}

Column\,10 presents the ratio between dust and photospheric
luminosity. This ratio can be used as a distance-independent mass-loss
indicator, which is not trivial, as it must be checked that the
variation of L$_{\rm{dust}}$/L$_{\star}$ over the sample is due to
different mass-loss rates and not only a difference in dust
composition, e.g. the oxides tend to generate much less emission than
 the magnesium-rich silicates, for a similar dust mass.  
We show our derived relation between L$_{\rm{dust}}$/L$_{\star}$ and
mass-loss rate, based on our modelling including the differences in
dust composition, in the left hand panel of
Fig.\,\ref{fig:spcorrelations}. When assuming an identical luminosity
for all stars, we find an almost one-to-one relation between mass-loss rate
and dust luminosity fraction. This shows that
L$_{\rm{dust}}$/L$_{\star}$ is a good mass-loss indicator. When correcting for distance and stellar
luminosity, the relation is less steep. This is a consequence of the
correlation between stellar luminosity and mass-loss rate discussed in
the next section. 

\subsection{A correlation between mass-loss rate and stellar luminosity}

The middle panel of Fig.\,\ref{fig:spcorrelations} shows a strong
correlation between mass-loss rate and stellar luminosity and therefore
also with stellar mass. The
best-fit linear relation is:
$$\dot{M} = 10^{-16.2 \pm 1.8} L_{\star}^{1.89\pm0.36}. $$
As an increase of the mass-loss rate with luminosity, i.e. size of the
object, is expected just from scaling arguments, we also show the dust luminosity fraction
as a function of stellar luminosity (right hand panel of
Fig.\,\ref{fig:spcorrelations} ). Interestingly, we find evidence for
a more efficient wind-driving mechanism in the more luminous stars. The
best-fit linear relation here is:
$$ L_{\rm{IR}}/L_{\star} = 10^{-7.5 \pm 1.7} L_{\star}^{1.06\pm0.30}. $$
We note also that we do not find a significant trend between mass-loss
rate and effective temperature.

\subsection{Dust composition as a function of mass-loss rate}

\begin{figure*}
\centering
  \resizebox{16cm}{12cm}{\includegraphics{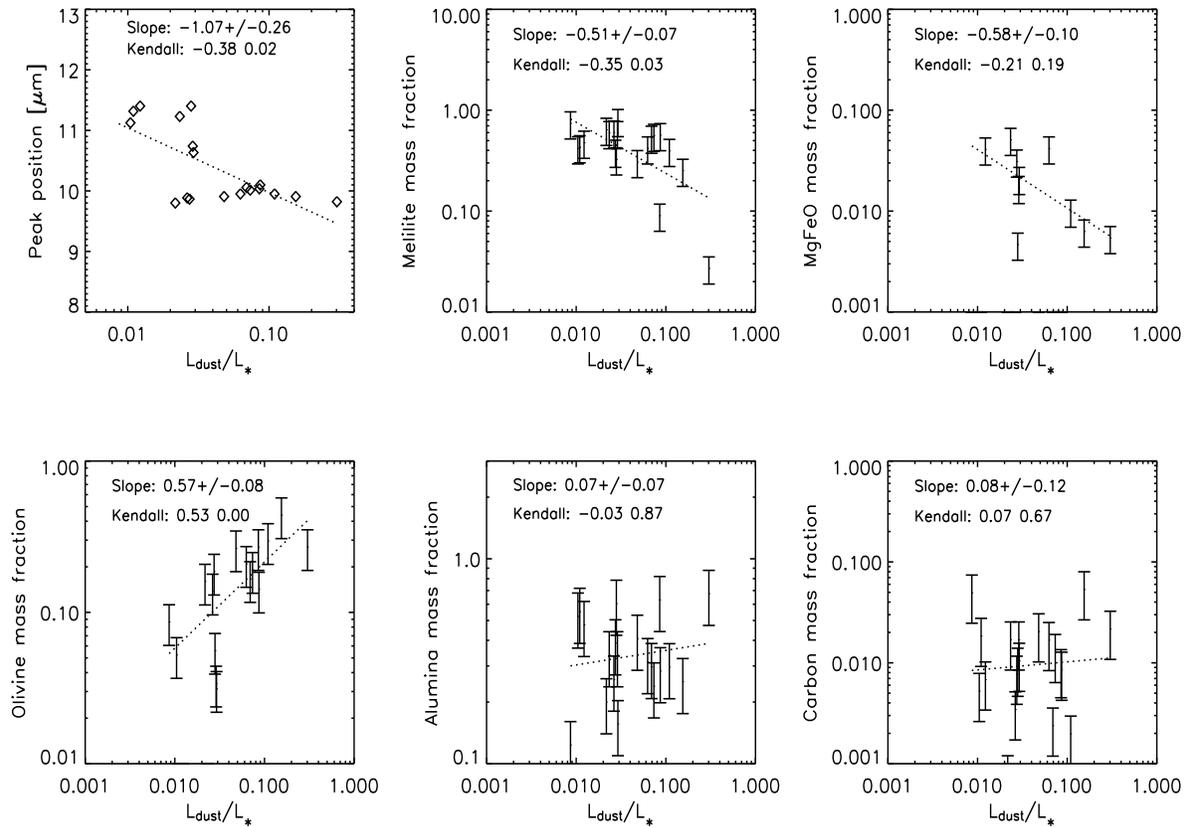}}
  \caption{Correlations between the different dust species and the
  mass-loss rate as traced by the dust luminosity (same analysis as in
  Fig.\,\ref{fig:spcorrelations}). } 
  \label{fig:comp_corr}
\end{figure*}

A purely empirical indication of the relation between mass-loss rate
and dust composition can be obtained by simply plotting peak position
of the 10-$\mu$m complex versus dust luminosity. This is done in
the upper left panel of Fig.\,\ref{fig:comp_corr}. With increasing mass-loss rate, the peak
position shifts from 11 to 10\,$\mu$m. For intermediate values
(e.g. BD+35\,4077, see top right in Fig.\,\ref{fig:fits}), it is difficult to assign a single peak
wavelength. 
%
%
%

 We search for correlations between dust composition and mass-loss rate
based on our modelling and find the following trends with increasing mass-loss rate:
\begin{enumerate}
\item the mass fraction of olivines increases
\item the melilite content decreases
\item the MgFeO content may decrease, but the Kendall test yields a
  low significance
\item the carbon and alumina content don't show a clear trend
\end{enumerate}

%

\begin{figure}
\centering
  \resizebox{\hsize}{!}{\includegraphics{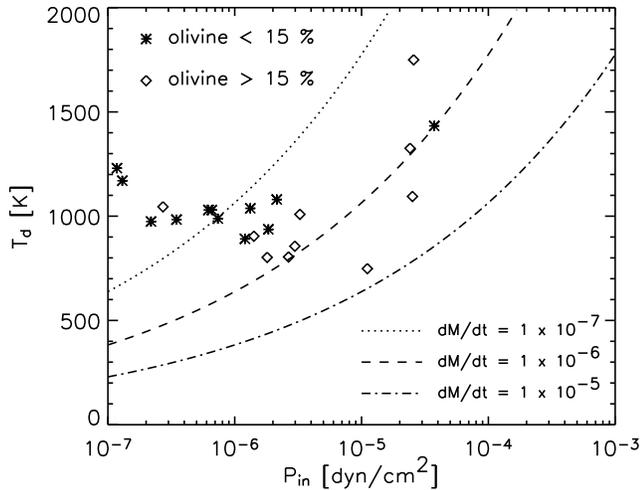}}
  \caption{Olivine content as a function of temperature and pressure
  at the inner edge of the dust shell. The lines represent regimes of
  constant mass-loss rate, as derived by \cite{Gail1986}.} 
  \label{fig:fig6}
\end{figure}

Since the freeze-out scenario actually predicts different compositions
for different temperature/pressure combinations in the wind, which of
course depend on the mass-loss rate, we show the olivine content as
a function of temperature and pressure at the inner edge of the dust
shell in Fig.\,\ref{fig:fig6}. It is clear from this figure that
indeed, larger fractions of olivine dust are formed at higher wind
densities as is predicted by the condensation sequence of
\cite{Tielens1990}, and as is seen in AGB stars with low mass-loss
rates \citep{Heras2005,Lebzelter2006}.

%

\section{Conclusions}
\label{sec:discussionandconclusions}
We have analyzed the ISO-SWS spectra of 27 red supergiant stars, out
of which 21 show significant mass loss through a dusty wind. The
determined properties of the winds around RSG share some important
characteristics with those around AGB stars with relatively low
mass-loss rates. 1) The dust has a high fraction of "simple" dust
species like metal-oxides. 2) The winds have a relatively low
abundance of silicates. 3) The fraction of silicates correlates well
with the mass-loss rate and/or the density and pressure at the base of
the wind, as predicted by the
canonical condensation sequence of \cite{Tielens1990}.

However, in some respects, RSG are not just the heavier twins of AGB
stars: 1) They show molecular bands only of di-atomic molecules (not
H$_2$O, CO$_2$ or SO$_2$). 2) The general slope of the SED from
near-IR to mid-IR wavelengths requires a source of continuous opacity
which, in the case of RSG, could be due to amorphous carbon.  3)
PAHs are observed in 4 sources, suggesting a strong influence by the
chromospheric radiation field.

Besides the correlation observed by \cite{Josselin2007} between
photospheric turbulence strength and mass-loss rate, we find that also
the stellar luminosity strenghtens the wind.

Although the correlations and detections presented in this paper are
significant (in a rough statistical sense) we are aware of possible 1)
degeneracies between the fitting parameters, 2) biases due to our
particular fitting strategy and model assumptions, and 3) biases due
to our limited set of dust constituents.

Major progress will be possible in the
near future with the 2nd generation VLTI instrumentation, in
particular MATISSE \citep{Lopez2006} will allow the
quasi-instanteneous imaging of many RSG from the near to mid-IR. This
will allow us to determine the actual spatial distribution of the
different dust components.

\acknowledgements{The authors would like to thank the anonymous
  referee for many valuable comments, and C.\,Kemper for a careful
  reading of the manuscript. KE gratefully acknowledges support from
  the Swedish Research Council.}

\bibliographystyle{aa}
\bibliography{references}

\begin{thebibliography}{77}
\expandafter\ifx\csname natexlab\endcsname\relax\def\natexlab#1{#1}\fi

\bibitem[{{Arenou} {et~al.}(1992){Arenou}, {Grenon}, \& {Gomez}}]{Arenou1992}
{Arenou}, F., {Grenon}, M., \& {Gomez}, A. 1992, \aap, 258, 104

\bibitem[{{Beck} {et~al.}(1992){Beck}, {Gail}, {Henkel}, \&
  {Sedlmayr}}]{Beck1992}
{Beck}, H.~K.~B., {Gail}, H.-P., {Henkel}, R., \& {Sedlmayr}, E. 1992, \aap,
  265, 626

\bibitem[{{Begemann} {et~al.}(1997){Begemann}, {Dorschner}, {Henning},
  {Mutschke}, {Guertler}, {Koempe}, \& {Nass}}]{Begemann1997}
{Begemann}, B., {Dorschner}, J., {Henning}, T., {et~al.} 1997, \apj, 476, 199

\bibitem[{{Bohren} \& {Huffman}(1983)}]{Bohren1983}
{Bohren}, C.~F. \& {Huffman}, D.~R. 1983, {Absorption and scattering of light
  by small particles} (New York: Wiley, 1983)

\bibitem[{{Bouwman}(2001)}]{Bouwman2001}
{Bouwman}, J. 2001, PhD thesis, University of Amsterdam

\bibitem[{{Bouwman} {et~al.}(2000){Bouwman}, {de Koter}, {van den Ancker}, \&
  {Waters}}]{Bouwman2000}
{Bouwman}, J., {de Koter}, A., {van den Ancker}, M.~E., \& {Waters},
  L.~B.~F.~M. 2000, \aap, 360, 213

\bibitem[{{Cami}(2002)}]{Cami2002}
{Cami}, J. 2002, PhD thesis, AA(University of Amsterdam)

\bibitem[{{Cardelli} {et~al.}(1989){Cardelli}, {Clayton}, \&
  {Mathis}}]{Cardelli1989}
{Cardelli}, J.~A., {Clayton}, G.~C., \& {Mathis}, J.~S. 1989, \apj, 345, 245

\bibitem[{{Chiar} \& {Tielens}(2006)}]{Chiar2006}
{Chiar}, J.~E. \& {Tielens}, A.~G.~G.~M. 2006, \apj, 637, 774

\bibitem[{{Cousins} \& {Lagerweij}(1971)}]{Cousins1971}
{Cousins}, A.~W.~J. \& {Lagerweij}, H.~C. 1971, Monthly Notes of the
  Astronomical Society of South Africa, 30, 12

\bibitem[{{de Graauw} {et~al.}(1996){de Graauw}, {Haser}, {Beintema},
  {Roelfsema}, {van Agthoven}, {Barl}, {Bauer}, {Bekenkamp}, {Boonstra},
  {Boxhoorn}, {Cote}, {de Groene}, {van Dijkhuizen}, {Drapatz}, {Evers},
  {Feuchtgruber}, {Frericks}, {Genzel}, {Haerendel}, {Heras}, {van der Hucht},
  {van der Hulst}, {Huygen}, {Jacobs}, {Jakob}, {Kamperman}, {Katterloher},
  {Kester}, {Kunze}, {Kussendrager}, {Lahuis}, {Lamers}, {Leech}, {van der
  Lei}, {van der Linden}, {Luinge}, {Lutz}, {Melzner}, {Morris}, {van Nguyen},
  {Ploeger}, {Price}, {Salama}, {Schaeidt}, {Sijm}, {Smoorenburg}, {Spakman},
  {Spoon}, {Steinmayer}, {Stoecker}, {Valentijn}, {Vandenbussche}, {Visser},
  {Waelkens}, {Waters}, {Wensink}, {Wesselius}, {Wiezorrek}, {Wieprecht},
  {Wijnbergen}, {Wildeman}, \& {Young}}]{deGraauw}
{de Graauw}, T., {Haser}, L.~N., {Beintema}, D.~A., {et~al.} 1996, \aap, 315,
  L49

\bibitem[{{Dijkstra} {et~al.}(2005){Dijkstra}, {Speck}, {Reid}, \&
  {Abraham}}]{Dijkstra2005}
{Dijkstra}, C., {Speck}, A.~K., {Reid}, R.~B., \& {Abraham}, P. 2005, \apjl,
  633, L133

\bibitem[{{Dorschner} {et~al.}(1995){Dorschner}, {Begemann}, {Henning},
  {Jaeger}, \& {Mutschke}}]{Dorschner1995}
{Dorschner}, J., {Begemann}, B., {Henning}, T., {Jaeger}, C., \& {Mutschke}, H.
  1995, \aap, 300, 503

\bibitem[{{Feautrier}(1964)}]{Feautrier1964}
{Feautrier}, P. 1964, C.R.Acad.Sc.Paris, 258, 3189

\bibitem[{{Gail} \& {Sedlmayr}(1986)}]{Gail1986}
{Gail}, H.-P. \& {Sedlmayr}, E. 1986, \aap, 166, 225

\bibitem[{{Gail} \& {Sedlmayr}(1999)}]{Gail1999}
{Gail}, H.-P. \& {Sedlmayr}, E. 1999, \aap, 347, 594

\bibitem[{{Gehrz}(1989)}]{Gehrz1989}
{Gehrz}, R. 1989, in IAU Symposium, Vol. 135, Interstellar Dust, ed. L.~J.
  {Allamandola} \& A.~G.~G.~M. {Tielens}, 445--+

\bibitem[{{Gielen} {et~al.}(2007){Gielen}, {van Winckel}, {Waters}, {Min}, \&
  {Dominik}}]{Gielen2007}
{Gielen}, C., {van Winckel}, H., {Waters}, L.~B.~F.~M., {Min}, M., \&
  {Dominik}, C. 2007, \aap, 475, 629

\bibitem[{{Gustafsson} {et~al.}(2008){Gustafsson}, {Edvardsson}, {Eriksson},
  {J{\o}rgensen}, {Nordlund}, \& {Plez}}]{Gustafsson2008}
{Gustafsson}, B., {Edvardsson}, B., {Eriksson}, K., {et~al.} 2008, \aap, 486,
  951

\bibitem[{{Harper} {et~al.}(2001){Harper}, {Brown}, \& {Lim}}]{Harper2001}
{Harper}, G.~M., {Brown}, A., \& {Lim}, J. 2001, \apj, 551, 1073

\bibitem[{{Harwit} {et~al.}(2001){Harwit}, {Malfait}, {Decin}, {Waelkens},
  {Feuchtgruber}, \& {Melnick}}]{Harwit2001}
{Harwit}, M., {Malfait}, K., {Decin}, L., {et~al.} 2001, \apj, 557, 844

\bibitem[{{Henning} {et~al.}(1995){Henning}, {Begemann}, {Mutschke}, \&
  {Dorschner}}]{Henning1995}
{Henning}, T., {Begemann}, B., {Mutschke}, H., \& {Dorschner}, J. 1995, \aaps,
  112, 143

\bibitem[{{Henning} \& {Stognienko}(1996)}]{Henning1996}
{Henning}, T. \& {Stognienko}, R. 1996, \aap, 311, 291

\bibitem[{{Heras} \& {Hony}(2005)}]{Heras2005}
{Heras}, A.~M. \& {Hony}, S. 2005, \aap, 439, 171

\bibitem[{{Herbst} \& {Assousa}(1977)}]{Herbst1977}
{Herbst}, W. \& {Assousa}, G.~E. 1977, \apj, 217, 473

\bibitem[{{Hoefner} {et~al.}(1998){Hoefner}, {Jorgensen}, {Loidl}, \&
  {Aringer}}]{Hoefner1998}
{Hoefner}, S., {Jorgensen}, U.~G., {Loidl}, R., \& {Aringer}, B. 1998, \aap,
  340, 497

\bibitem[{{H{\"o}fner}(2008)}]{Hoefner2008}
{H{\"o}fner}, S. 2008, \aap, 491, L1

\bibitem[{{H{\"o}fner} \& {Andersen}(2007)}]{Hoefner2007}
{H{\"o}fner}, S. \& {Andersen}, A.~C. 2007, \aap, 465, L39

\bibitem[{{Hony} {et~al.}(2001){Hony}, {Van Kerckhoven}, {Peeters}, {Tielens},
  {Hudgins}, \& {Allamandola}}]{Hony2001}
{Hony}, S., {Van Kerckhoven}, C., {Peeters}, E., {et~al.} 2001, \aap, 370, 1030

\bibitem[{{Humphreys}(1978)}]{Humphreys1978}
{Humphreys}, R.~M. 1978, \apjs, 38, 309

\bibitem[{{Humphreys} \& {Ney}(1974)}]{Humphreys1974}
{Humphreys}, R.~M. \& {Ney}, E.~P. 1974, \apj, 194, 623

\bibitem[{{Jaeger} {et~al.}(1994){Jaeger}, {Mutschke}, {Begemann}, {Dorschner},
  \& {Henning}}]{Jaeger1994}
{Jaeger}, C., {Mutschke}, H., {Begemann}, B., {Dorschner}, J., \& {Henning}, T.
  1994, \aap, 292, 641

\bibitem[{{Jennings} \& {Sada}(1998)}]{Jennings1998}
{Jennings}, D.~E. \& {Sada}, P.~V. 1998, Science, 279, 844

\bibitem[{{Johnson} {et~al.}(1966{\natexlab{a}}){Johnson}, {Iriarte},
  {Mitchell}, \& {Wisniewskj}}]{Johnson1966a}
{Johnson}, H.~L., {Iriarte}, B., {Mitchell}, R.~I., \& {Wisniewskj}, W.~Z.
  1966{\natexlab{a}}, Communications of the Lunar and Planetary Laboratory, 4,
  99

\bibitem[{{Johnson} {et~al.}(1966{\natexlab{b}}){Johnson}, {Mendoza V.}, \&
  {Eugenio}}]{Johnson1966b}
{Johnson}, H.~L., {Mendoza V.}, \& {Eugenio}, E. 1966{\natexlab{b}}, Annales
  d'Astrophysique, 29, 525

\bibitem[{{Josselin} \& {Plez}(2007)}]{Josselin2007}
{Josselin}, E. \& {Plez}, B. 2007, \aap, 469, 671

\bibitem[{{Kemper} {et~al.}(2002){Kemper}, {de Koter}, {Waters}, {Bouwman}, \&
  {Tielens}}]{Kemper2002}
{Kemper}, F., {de Koter}, A., {Waters}, L.~B.~F.~M., {Bouwman}, J., \&
  {Tielens}, A.~G.~G.~M. 2002, \aap, 384, 585

\bibitem[{{Kessler} {et~al.}(1996){Kessler}, {Steinz}, {Anderegg}, {Clavel},
  {Drechsel}, {Estaria}, {Faelker}, {Riedinger}, {Robson}, {Taylor}, \&
  {Xim{\'e}nez de Ferr{\'a}n}}]{Kessler1996}
{Kessler}, M.~F., {Steinz}, J.~A., {Anderegg}, M.~E., {et~al.} 1996, \aap, 315,
  L27

\bibitem[{{Kharchenko}(2001)}]{Kharchenko2001}
{Kharchenko}, N.~V. 2001, Kinematika i Fizika Nebesnykh Tel, 17, 409

\bibitem[{{Kholopov} {et~al.}(1998){Kholopov}, {Samus}, {Frolov}, {Goranskij},
  {Gorynya}, {Karitskaya}, {Kazarovets}, {Kireeva}, {Kukarkina}, {Kurochkin},
  {Medvedeva}, {Pastukhova}, {Perova}, {Rastorguev}, \&
  {Shugarov}}]{Kholopov1998}
{Kholopov}, P.~N., {Samus}, N.~N., {Frolov}, M.~S., {et~al.} 1998, in Combined
  General Catalogue of Variable Stars, 4.1 Ed (II/214A). (1998), 0--+

\bibitem[{{Koike} {et~al.}(1995){Koike}, {Kaito}, {Yamamoto}, {Shibai},
  {Kimura}, \& {Suto}}]{Koike1995}
{Koike}, C., {Kaito}, C., {Yamamoto}, T., {et~al.} 1995, Icarus, 114, 203

\bibitem[{{Kraemer} {et~al.}(2002){Kraemer}, {Sloan}, {Price}, \&
  {Walker}}]{Kraemer2002}
{Kraemer}, K.~E., {Sloan}, G.~C., {Price}, S.~D., \& {Walker}, H.~J. 2002,
  \apjs, 140, 389

\bibitem[{{Lebzelter} {et~al.}(2006){Lebzelter}, {Posch}, {Hinkle}, {Wood}, \&
  {Bouwman}}]{Lebzelter2006}
{Lebzelter}, T., {Posch}, T., {Hinkle}, K., {Wood}, P.~R., \& {Bouwman}, J.
  2006, \apjl, 653, L145

\bibitem[{{Lee}(1970)}]{Lee1970}
{Lee}, T.~A. 1970, \apj, 162, 217

\bibitem[{{Lejeune} \& {Schaerer}(2001)}]{Lejeune2001}
{Lejeune}, T. \& {Schaerer}, D. 2001, \aap, 366, 538

\bibitem[{{Levesque} {et~al.}(2005){Levesque}, {Massey}, {Olsen}, {Plez},
  {Josselin}, {Maeder}, \& {Meynet}}]{Levesque2005}
{Levesque}, E.~M., {Massey}, P., {Olsen}, K.~A.~G., {et~al.} 2005, \apj, 628,
  973

\bibitem[{{Lopez} {et~al.}(2006){Lopez}, {Wolf}, {Lagarde}, {Abraham},
  {Antonelli}, {Augereau}, {Beckman}, {Behrend}, {Berruyer}, {Bresson},
  {Chesneau}, {Clausse}, {Connot}, {Demyk}, {Danchi}, {Dugu{\'e}}, {Flament},
  {Glazenborg}, {Graser}, {Henning}, {Hofmann}, {Heininger}, {Hugues}, {Jaffe},
  {Jankov}, {Kraus}, {Laun}, {Leinert}, {Linz}, {Mathias}, {Meisenheimer},
  {Matter}, {Menut}, {Millour}, {Neumann}, {Nussbaum}, {Niedzielski},
  {Mosonic}, {Petrov}, {Ratzka}, {Robbe-Dubois}, {Roussel}, {Schertl},
  {Schmider}, {Stecklum}, {Thiebaut}, {Vakili}, {Wagner}, {Waters}, \&
  {Weigelt}}]{Lopez2006}
{Lopez}, B., {Wolf}, S., {Lagarde}, S., {et~al.} 2006, in Presented at the
  Society of Photo-Optical Instrumentation Engineers (SPIE) Conference, Vol.
  6268, Advances in Stellar Interferometry. Edited by Monnier, John D.;
  Sch{\"o}ller, Markus; Danchi, William C.. Proceedings of the SPIE, Volume
  6268, pp. 62680Z (2006).

\bibitem[{{Massey}(2003)}]{Massey2003}
{Massey}, P. 2003, \araa, 41, 15

\bibitem[{{Massey} {et~al.}(2005){Massey}, {Plez}, {Levesque}, {Olsen},
  {Clayton}, \& {Josselin}}]{Massey2005}
{Massey}, P., {Plez}, B., {Levesque}, E.~M., {et~al.} 2005, \apj, 634, 1286

\bibitem[{{Mathis} {et~al.}(1977){Mathis}, {Rumpl}, \&
  {Nordsieck}}]{Mathis1977}
{Mathis}, J.~S., {Rumpl}, W., \& {Nordsieck}, K.~H. 1977, \apj, 217, 425

\bibitem[{{Matsuura} {et~al.}(2005){Matsuura}, {Zijlstra}, {van Loon},
  {Yamamura}, {Markwick}, {Whitelock}, {Woods}, {Marshall}, {Feast}, \&
  {Waters}}]{Matsuura2005}
{Matsuura}, M., {Zijlstra}, A.~A., {van Loon}, J.~T., {et~al.} 2005, \aap, 434,
  691

\bibitem[{{Mendoza}(1967)}]{Mendoza1967}
{Mendoza}, E.~E. 1967, Boletin de los Observatorios Tonantzintla y Tacubaya, 4,
  149

\bibitem[{{Mihalas}(1978)}]{Mihalas1978}
{Mihalas}, D. 1978, {Stellar atmospheres /2nd edition/} (San Francisco,
  W.~H.~Freeman and Co., 1978.~650 p.)

\bibitem[{{Min} {et~al.}(2005){Min}, {Hovenier}, \& {de Koter}}]{Min2005}
{Min}, M., {Hovenier}, J.~W., \& {de Koter}, A. 2005, \aap, 432, 909

\bibitem[{{Mutschke} {et~al.}(1998){Mutschke}, {Begemann}, {Dorschner},
  {Guertler}, {Gustafson}, {Henning}, \& {Stognienko}}]{Mutschke1998}
{Mutschke}, H., {Begemann}, B., {Dorschner}, J., {et~al.} 1998, \aap, 333, 188

\bibitem[{{Neugebauer} {et~al.}(1984){Neugebauer}, {Habing}, {van Duinen},
  {Aumann}, {Baud}, {Beichman}, {Beintema}, {Boggess}, {Clegg}, {de Jong},
  {Emerson}, {Gautier}, {Gillett}, {Harris}, {Hauser}, {Houck}, {Jennings},
  {Low}, {Marsden}, {Miley}, {Olnon}, {Pottasch}, {Raimond}, {Rowan-Robinson},
  {Soifer}, {Walker}, {Wesselius}, \& {Young}}]{Neugebauer1984}
{Neugebauer}, G., {Habing}, H.~J., {van Duinen}, R., {et~al.} 1984, \apjl, 278,
  L1

\bibitem[{{Nicolet}(1978)}]{Nicolet1978}
{Nicolet}, B. 1978, \aaps, 34, 1

\bibitem[{{Ohnaka}(2004)}]{Ohnaka2004}
{Ohnaka}, K. 2004, \aap, 424, 1011

\bibitem[{{Peeters}(2002)}]{Peeters2002}
{Peeters}, E. 2002, PhD thesis, Proefschrift, Rijksuniversiteit Groningen,
  2002, 237 p.

\bibitem[{{Perrin} {et~al.}(2004){Perrin}, {Ridgway}, {Mennesson}, {Cotton},
  {Woillez}, {Verhoelst}, {Schuller}, {Coud{\'e} du Foresto}, {Traub},
  {Millan-Gabet}, \& {Lacasse}}]{Perrin2004}
{Perrin}, G., {Ridgway}, S.~T., {Mennesson}, B., {et~al.} 2004, \aap, 426, 279

\bibitem[{{Perrin} {et~al.}(2005){Perrin}, {Ridgway}, {Verhoelst}, {Schuller},
  {Coud{\'e} Du Foresto}, {Traub}, {Millan-Gabet}, \& {Lacasse}}]{Perrin2005}
{Perrin}, G., {Ridgway}, S.~T., {Verhoelst}, T., {et~al.} 2005, \aap, 436, 317

\bibitem[{{Perrin} {et~al.}(2007){Perrin}, {Verhoelst}, {Ridgway}, {Cami}, {Nhu
  Nguyen}, {Chesneau}, {Lopez}, {Leinert}, \& {Richichi}}]{Perrin2007}
{Perrin}, G., {Verhoelst}, T., {Ridgway}, S.~T., {et~al.} 2007, ArXiv e-prints,
  709

\bibitem[{{Perryman} {et~al.}(1997){Perryman}, {Lindegren}, {Kovalevsky},
  {Hoeg}, {Bastian}, {Bernacca}, {Cr{\'e}z{\'e}}, {Donati}, {Grenon}, {van
  Leeuwen}, {van der Marel}, {Mignard}, {Murray}, {Le Poole}, {Schrijver},
  {Turon}, {Arenou}, {Froeschl{\'e}}, \& {Petersen}}]{Perryman1997}
{Perryman}, M.~A.~C., {Lindegren}, L., {Kovalevsky}, J., {et~al.} 1997, \aap,
  323, L49

\bibitem[{{Preibisch} {et~al.}(1993){Preibisch}, {Ossenkopf}, {Yorke}, \&
  {Henning}}]{Preibisch1993}
{Preibisch}, T., {Ossenkopf}, V., {Yorke}, H.~W., \& {Henning}, T. 1993, \aap,
  279, 577

\bibitem[{{Ryde} {et~al.}(2006){Ryde}, {Richter}, {Harper}, {Eriksson}, \&
  {Lambert}}]{Ryde2006}
{Ryde}, N., {Richter}, M.~J., {Harper}, G.~M., {Eriksson}, K., \& {Lambert},
  D.~L. 2006, \apj, 645, 652

\bibitem[{{Skrutskie} {et~al.}(2006){Skrutskie}, {Cutri}, {Stiening},
  {Weinberg}, {Schneider}, {Carpenter}, {Beichman}, {Capps}, {Chester},
  {Elias}, {Huchra}, {Liebert}, {Lonsdale}, {Monet}, {Price}, {Seitzer},
  {Jarrett}, {Kirkpatrick}, {Gizis}, {Howard}, {Evans}, {Fowler}, {Fullmer},
  {Hurt}, {Light}, {Kopan}, {Marsh}, {McCallon}, {Tam}, {Van Dyk}, \&
  {Wheelock}}]{Skrutskie2006}
{Skrutskie}, M.~F., {Cutri}, R.~M., {Stiening}, R., {et~al.} 2006, \aj, 131,
  1163

\bibitem[{{Sloan} {et~al.}(2003){Sloan}, {Kraemer}, {Price}, \&
  {Shipman}}]{Sloan2003}
{Sloan}, G.~C., {Kraemer}, K.~E., {Price}, S.~D., \& {Shipman}, R.~F. 2003,
  \apjs, 147, 379

\bibitem[{{Speck} {et~al.}(2000){Speck}, {Barlow}, {Sylvester}, \&
  {Hofmeister}}]{Speck2000}
{Speck}, A.~K., {Barlow}, M.~J., {Sylvester}, R.~J., \& {Hofmeister}, A.~M.
  2000, \aaps, 146, 437

\bibitem[{{Sylvester} {et~al.}(1994){Sylvester}, {Barlow}, \&
  {Skinner}}]{Sylvester1994}
{Sylvester}, R.~J., {Barlow}, M.~J., \& {Skinner}, C.~J. 1994, \mnras, 266, 640

\bibitem[{{Sylvester} {et~al.}(1998){Sylvester}, {Skinner}, \&
  {Barlow}}]{Sylvester1998}
{Sylvester}, R.~J., {Skinner}, C.~J., \& {Barlow}, M.~J. 1998, \mnras, 301,
  1083

\bibitem[{{Tielens}(1990)}]{Tielens1990}
{Tielens}, A.~G.~G.~M. 1990, in From Miras to Planetary Nebulae: Which Path for
  Stellar Evolution?, ed. M.~O. {Mennessier} \& A.~{Omont}, 186--200

\bibitem[{{Tsuji}(2000)}]{Tsuji2000}
{Tsuji}, T. 2000, \apj, 538, 801

\bibitem[{{van Boekel} {et~al.}(2005){van Boekel}, {Min}, {Waters}, {de Koter},
  {Dominik}, {van den Ancker}, \& {Bouwman}}]{vanBoekel2005}
{van Boekel}, R., {Min}, M., {Waters}, L.~B.~F.~M., {et~al.} 2005, \aap, 437,
  189

\bibitem[{{Verhoelst} {et~al.}(2006){Verhoelst}, {Decin}, {van Malderen},
  {Hony}, {Cami}, {Eriksson}, {Perrin}, {Deroo}, {Vandenbussche}, \&
  {Waters}}]{Verhoelst2006}
{Verhoelst}, T., {Decin}, L., {van Malderen}, R., {et~al.} 2006, \aap, 447, 311

\bibitem[{{Wawrukiewicz} \& {Lee}(1974)}]{Wawrukiewicz1974}
{Wawrukiewicz}, A.~S. \& {Lee}, T.~A. 1974, \pasp, 86, 51

\bibitem[{{Woitke}(2006)}]{Woitke2006}
{Woitke}, P. 2006, \aap, 460, L9

\bibitem[{{Woitke} \& {Niccolini}(2005)}]{Woitke2005}
{Woitke}, P. \& {Niccolini}, G. 2005, \aap, 433, 1101

\end{thebibliography}

\end{document}